\documentclass[aps,pra,groupaddress]{revtex4-1}
\usepackage{graphicx}
\usepackage{amsmath,amssymb}
\usepackage{bm}
\usepackage{tikz}
\usetikzlibrary{positioning}

\newcommand{\ee}[1]{\relax\ifmmode 10^{#1} \else 10$^{#1}$\fi}

\renewcommand{\eqref}[1]{Eq.~(\ref{#1})}

\newcommand{\TtwoB}{T^\mathrm{2B}}

\newcommand{\ha}{\hat{a}}
\newcommand{\had}{\hat{a}{}^\dagger}

\newcommand{\AD}{A_\mathrm{D}}
\newcommand{\AX}{A_\mathrm{X}}
\newcommand{\omgD}{\omega_\mathrm{D}}
\newcommand{\omgX}{\omega_\mathrm{X}}

\begin{document}
\title{On the long-term stability of space-time crystals}

\author{J. Smits${}^1$}
\author{H.T.C. Stoof${}^2$}
\author{P. van der Straten${}^1$}
\email[]{p.vanderstraten@uu.nl}

\affiliation{${}^1$Debye Institute for Nanomaterials and Center for Extreme Matter and Emergent Phenomena, Utrecht University, PO Box 80.000, 3508 TA Utrecht,The Netherlands\\
${}^2$Institute for Theoretical Physics and Center for Extreme Matter and Emergent Phenomena, Utrecht University, PO Box 80.000, 3508 TA Utrecht,The Netherlands}
\date{\today}%

\begin{abstract}
We investigate a space-time crystal in a superfluid Bose gas. Using a well-controlled periodic drive we excite only one crystalline mode in the system, which can be accurately modeled in the rotating frame of the drive. Using holographic imaging we observe the stability of the crystal over an extended period of time and show the robustness of its structure in both space and time. By introducing a fourth-order term in the Hamiltonian we show that the crystal stabilizes at a fixed number of quanta. The results of the model are compared to the experimental data and show good agreement, with a small number of free parameters. The results yield insights in the long-term stability of the crystal, which can only be obtained by the combination of the extended control in the experiment and the nearly {\it ab-initio} character of the model. From the model we derive a phase diagram of the system, which can be exploited in the future to study the phase transitions for this new state of matter in even more detail. 
\end{abstract}

\maketitle

\section{Introduction}
The concept of time crystals was first introduced by Frank Wilczek in 2012 as a novel state of matter, breaking the continuous time translation symmetry in the ground state  \cite{classtxtal,quanttxtal}. This sparked intense discussion on the feasibility of time crystals \cite{bruno13,nozieres13}, leading to a no-go proof pertaining to breaking of continuous time translation symmetry in a quantum-mechanical ground state \cite{watanabe15}. Therefore, the focus shifted from breaking the continuous symmetry to breaking a discrete one, introducing the Floquet time crystal. Many proposals have since followed to realize such a Floquet time crystal in different systems \cite{guo13,sacha15,khemani16,keyserlingk16,else16,else17,russomanno17,zeng17,gong18,iemini18,mizuta18,gambetta19,cosme19,heugel19,giergiel20,kuros20} and several experimental realizations using different systems have been demonstrated \cite{zhang17,choi17,autti18,rovny18_lett,rovny18_b,smits18,trager19}.

In a previous work we presented the observation of a space-time crystal in a cigar-shaped Bose-Einstein condensate (BEC) \cite{smits18}. By modulating the trap in the radial direction a radial breathing mode is excited. This radial breathing mode in turn acts as the driving force for a collection of axial modes of the cloud, realizing a Floquet time crystal which also exhibits high spatial ordering. In that study it was hard to conclusively show long term stability due to the spurious excitation of additional collective modes and loss of particles due to imaging. In this paper we will show how we have overcome these problems and report on our results. 


The paper is organized as follows. In Sec.~\ref{sc:experiment} the experimental setup is discussed, with an emphasis on the improvements to the experimental setup and sequence compared to the previous work. In Sec.~\ref{sc:modelling} the model of Ref. \cite{liao19} is expanded with a fourth-order interaction term, which provides insight into the stabilization mechanism bounding the growth of the observed pattern. In Sec.~\ref{sc:breaking} we explicitly show the broken time-symmetry in our experiment. The updated model is used in Sec.~\ref{sc:growth} to relate the growth of the crystal in the experiment to the driving strength, detuning and the value of the fourth-order term. Finally, in Sec.~\ref{sc:stability} the saturated amplitude of the space-time crystal for different driving strengths is compared to the model.


\section{Experiment\label{sc:experiment}}
In our experiment a superfluid cloud of sodium atoms is trapped in an elongated magnetic trap, resulting in a cigar-shaped Bose-Einstein condensate. By modulating the trap in the radial direction, a radial size oscillation is induced. In previous work \cite{smits18} we have demonstrated that this mode, the radial breathing mode, couples to a higher-order axial mode through a non-linear coupling. The induced higher-order axial mode is shown to satisfy all the requirements of a time crystal, with the addition of also exhibiting spatial ordering.

A bottleneck in previous experiments was the loss of particles as a result of photon scattering during imaging. For this work an imaging method based on off-axis holography was used~\cite{kadlecek01,smits20} instead of a phase-contrast imaging method. As a result, the BEC is imaged at comparable signal-to-noise ratio (SNR), but at loss rates a factor of 50 lower than in previous experiments. Bose-Einstein condensates are recorded with a sampling rate of up to 1 kHz, and up to 250 images are taken of a single sample with an approximate total particle loss of less than 20\% during the entire imaging sequence.

Further improvements are made to the method by which the sample is prepared and the initial modulation is applied. As the resonance conditions for each axial mode depend on the axial extent (length) of the BEC, the length of the cloud has to be constant to draw conclusions about true single mode long-term stability. In this work, the magnetic trap containing the atom cloud is deformed to the parameters which will be used in the experiment prior to crossing the transition to a Bose-Einstein condensate. Any excitations due to the trap deformation will damp out and thermalize or be quelled by the subsequent cooling process. After the BEC is created, the radial confinement is modulated using a triangular waveform with a period of 10\,ms, which is approximately the time-scale associated with the radial trapping frequency, and varying modulation depth. The short driving period and moderate modulation depth ensure the relative amplitude of the oscillation of the length is always less than 2.5\%.

\begin{figure}[h]
\includegraphics{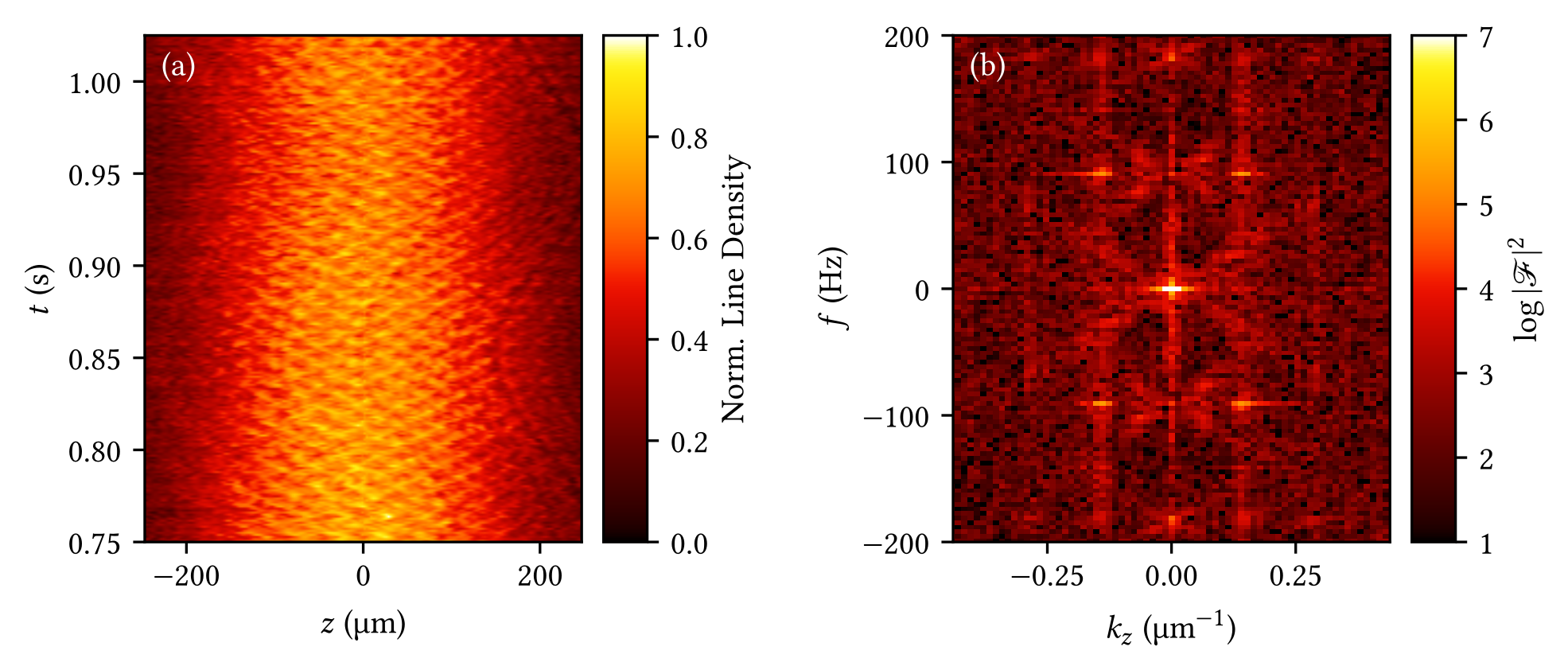}
\caption{(Color online). (a) Space-time crystal over an extended range of time. The signal strength remains nearly constant over the whole range. The ``wavy" pattern is due to center-of-mass motion correction and inhomogeneity of the camera sensor sensitivity. (b) Fourier spectrum of the space-time crystal on a logarithmic scale. \label{real_space_time}}
\end{figure}

The experiments in this work are performed in a cylindrically symmetric magnetic trap with trapping parameters $(\omega_\rho,\omega_z) = 2\pi\times(92.09(2),5.048(10))\,\mathrm{Hz}$ with about  $10^7$ particles at a condensate fraction of above 90\%. In the experiment the radial breathing mode, which functions as the periodic driving mode, is excited and the sample is left to evolve freely for a variable time after the initial excitation. After this hold time, a sequence of 250 consecutive images is recorded with a 1.1\,ms interval. At this imaging interval we fully resolve the radial breathing mode, which has the highest frequency of the excited collective modes in the system. Each image yields a two-dimensional column-density profile, with one of the radial directions integrated over. After correcting only for displacement and tilt of the sample, the remaining radial direction is integrated out and a line-density profile along the axial direction is constructed for each frame. By collecting all line-density profiles in a two-dimensional (space-time) density plot, the time behavior of the axial modes is visualized (Fig. \ref{real_space_time}a). With the used sampling rate and the number of images taken, we are able to resolve $\approx30$ periods of oscillation of the time crystal in a single measurement. Over the entire measurement run a moderate loss of signal is observed corresponding to particle loss due to imaging. The oscillation of the length of the condensate is in this case less than 1\%.

A two-dimensional Fourier transform of the space-time resolved data is shown in Fig. \ref{real_space_time}b. A rectangle in $(f,k)$ space is observed, the clear signature of space-time crystalline ordering, in analogy with well-known X-ray Bragg spectroscopy. Peaks related to the space-time crystal appear at $f = \pm90.5\,\textrm{Hz}$, $k_z = \pm0.14\,\mu\textrm{m}^{-1}$. The width of a frequency bin is approximately 3\,Hz. No broadening is observed along the frequency axis, implying a very well-defined frequency. Indeed, direct fitting of the driving mode and crystal yield $\omgD=2\pi\times 183.26(4)\,\textrm{Hz}$ and $\omgX = 2\pi\times 91.61(3)\,\textrm{Hz}$, respectively, confirming that within the quoted uncertainty the space-time crystal oscillates with half the drive frequency. Some broadening along the wavevector axis observed for the four peaks associated with the crystalline order is attributed to the inhomogeneous spatial profile, which in turn is a result of the inhomogeneous trap shape. The peak at the center of the Fourier space corresponds to the equilibrium profile and is broadened along the spatial (wavenumber) axis due to the spatial profile of the BEC and along the frequency axis due to particle losses caused by the imaging.



\section{Modelling \label{sc:modelling}}
\begin{figure}[h]
\includegraphics[]{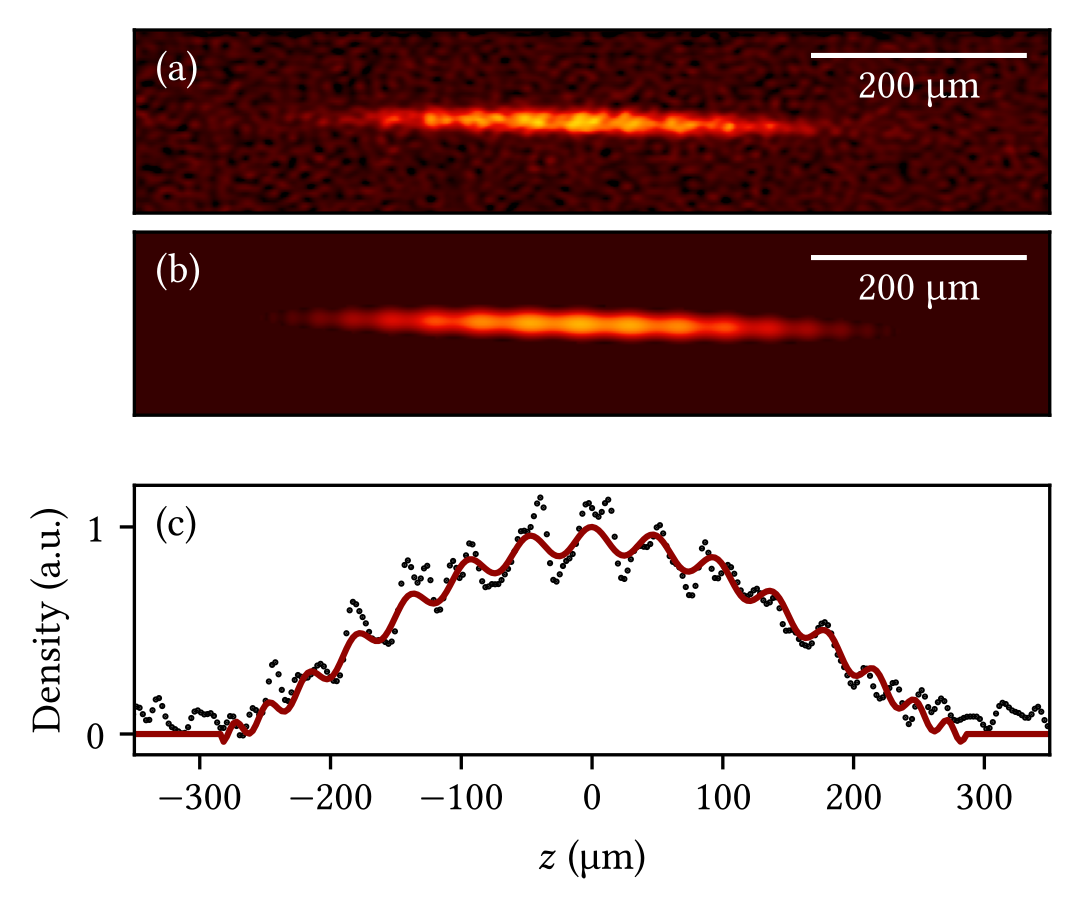}
\caption{A comparison of model and experimental observation of the column density profile.  Full column density profile  as observed in the experiment \emph{(a)} and a two-dimension fit on the data \emph{(b)} using the model in Ref. \cite{liao19}. \emph{(c)} A cut through the center of the cloud. The points represent data, the solid line is a fit of the model for $j=40$.
\label{profile}}
\end{figure}

The excitation observed in the experiment is quantified by fitting the column density profile with the sum of a Thomas-Fermi density profile and a higher-order pattern along the length of the cloud, where the latter is defined as 
\begin{equation}
    L_j(\bar{z}) = P_{j+2}(\bar{z}) - P_j(\bar{z}),
    \label{eq:modeprofile}
\end{equation}
where $P_j$ represent Legendre polynomials and $\bar{z}$ is the coordinate along the axial direction, normalized to the length of, and centered on, the cloud. Note that the indexing $j$ has been changed compared to previous work \cite{liao19} to include both even and odd $L_j(z)$ instead of only even modes. In the current work, $j$ alternated between $j=40$ and $j=41$ without clear relation to the particle number and condensate length. A fit of this model to a two-dimensional column density profile can be seen in Figure \ref{profile}.

\begin{figure}[h]
\includegraphics[]{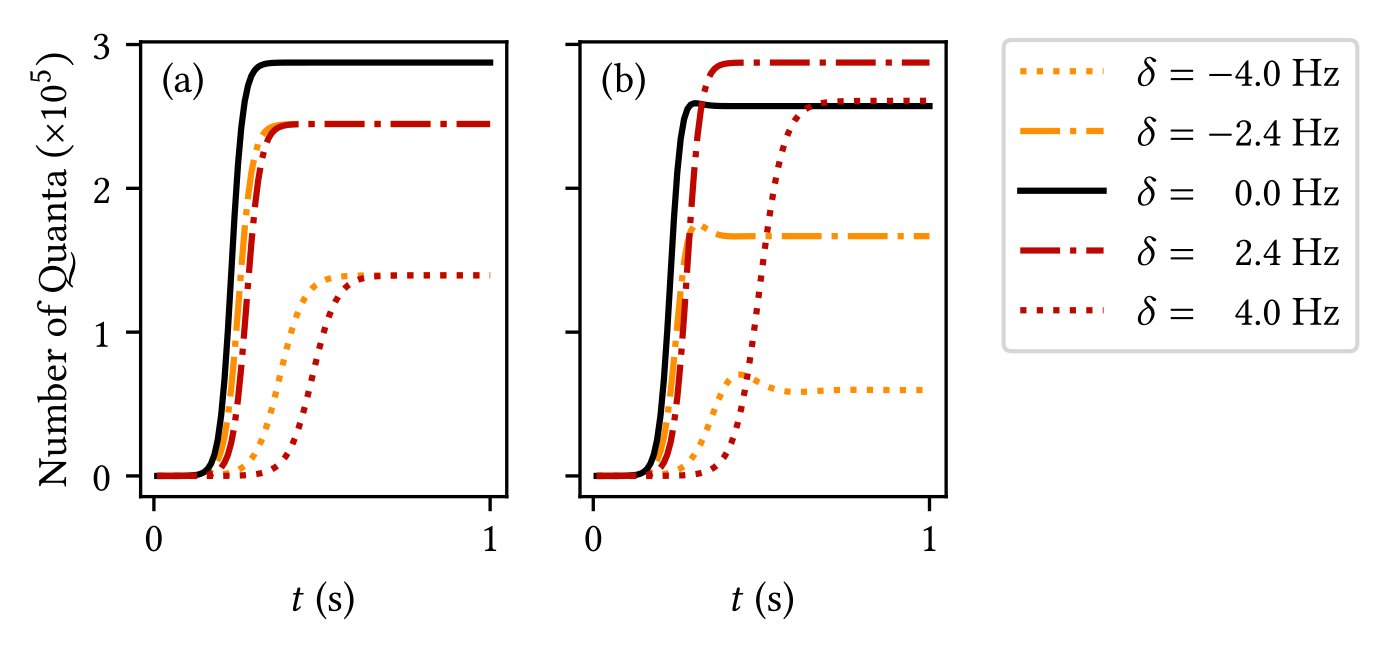}
\caption{Number of quanta as a function of time for $\omgD = 2\pi\times 183\,\textrm{Hz}$, $g'' = 10^{-4}\,\textrm{s}^{-1}$, $\AD=0.1$ and \emph{(a)} $g' = 0\,\textrm{s}^{-1}$ or \emph{(b)} $g' = 10^{-4}\,\textrm{s}^{-1}$. The dynamics are simulated for up to a second assuming $a=1$ at $t=0$. The value $|a|^2$ corresponds to the number of quanta in the crystalline mode. \label{model_sim}}
\end{figure}

The dynamics of the space-time crystal were modeled in Ref. \cite{liao19} up to third order. The model describes the space-time crystal as a single periodic driving mode in a collection of axial modes of the BEC. This model can be reduced, for only one dominant axial mode in the rotating frame of the drive, to%
\begin{equation}
    \hat{H} = -\hbar \delta \had\ha + \frac{\hbar\omgD\AD}{8} ( \had\had + \ha\ha ) ,
    \label{eq:lei_2ndorder}
\end{equation}%
where $\delta$ is the detuning from resonance, $\omgD$ is the driving frequency, $\AD$ is the relative driving amplitude and $\ha{}^{(\dagger)}$ is the annihilation (creation) operator of a quantum in the dominant axial mode. This model predicts infinite growth as long as $\AD > 4|\delta|/\omgD$. However, in the experiment we do not observe infinite growth for the space-time crystal but rather stabilization at some finite amplitude. We therefore expand the model with the only fourth-order term that is resonant in the rotating frame, resulting in%
\begin{equation}
    \hat{H} = -\hbar \delta \had\ha + \frac{\hbar\omgD\AD}{8} ( \had\had + \ha\ha ) + \frac{\hbar g}{2} \had\had\ha\ha ,
    \label{eq:lei_4thorder}
\end{equation}%
where $g=g'+i g''$ is a complex-valued parameter capturing all the fourth-order terms not yet calculated in earlier work \cite{liao19}, but which are introduced here in the model. If $\langle\ha\rangle\gg1$, the creation (annihilation) operators can be reinterpreted as classical fields, $\hat{a}^{(\dagger)} \rightarrow a^{(*)}$, and the equations of motion for the fields $a$ and $a^*$ are given by
\begin{align}
    i \frac{\textrm{d}}{\textrm{d}t} a &= \left( -\delta + g|a|^2 \right) a + \frac{\omgD\AD}{4} a^* \label{eq:a_eom},\\
    -i\frac{\textrm{d}}{\textrm{d}t} a^* &= \left( -\delta + g^*|a|^2 \right) a^* + \frac{ \omgD\AD}{4} a \label{eq:as_eom}.
\end{align}
Note that $|a|^2$ is then equivalent to the number operator $\had\ha$ and represents the number of quanta in the dominant axial mode, and growth is now bounded by the imaginary part of $g$. In Fig.~\ref{model_sim} the equations of motion are solved for different starting conditions. Due to the exponential growth, it takes 100--150\,ms for the pattern to be observable. It is seen that when the real part of $g$ is zero, the final value of $|a|^2$ only relies on the magnitude of the detuning $\delta$, but the growth rate also relies on the sign of $\delta$. For non-zero real part of $g$ the final value of $|a|^2$ relies on both sign and magnitude of the detuning. The maximum value for the number of quanta $|a|^2$ is shifted towards positive detuning for positive $g'$.

To further elucidate the behavior of the equation of motion, we can write the evolution of $|a|$  from Eqs. (\ref{eq:a_eom}) and (\ref{eq:as_eom}) close to equilibrium in terms of an effective potential $V(|a|)$ given by
\begin{equation}
    \frac{\textrm{d}^2}{\textrm{d}t^2} |a| = - \frac{\textrm{d}}{\textrm{d}|a|}V(|a|)\qquad\textrm{with}\qquad V(|a|)=-\frac{1}{2}\left[ \left( \frac{\omgD\AD}{4} \right)^2 - \delta^2 \right] |a|^2 -\frac{g'}{2}\delta|a|^4+\frac{|g|^2}{6}|a|^6.%
    \label{eq:pot_absa}
\end{equation}%
Interestingly, this effective potential is equivalent to a Landau free energy for a system with a tricritical point at $(\AD,\delta) = (0,0)$~\cite{boek_stoof}. The extreme values of the potential are found by calculating $\frac{\textrm{d}}{\textrm{d}|a|} V(|a|) = 0$ and are given, for $g\neq0$, by 
\begin{equation}
    |a|^2 = \frac{g'}{|g|^2}  \delta + \frac{1}{|g|}\sqrt{   \left(\frac{\hbar \omgD\AD}{4} \right)^2 - \frac{(g'')^2}{|g|^2} \delta^2}.
    \label{eq:eq_absa}
\end{equation}%

Note that only $g$ with $g''<0$ result in stable solutions. For purely imaginary $g$, in case $g''>0$ super-exponential growth is observed, while $g''=0$ results in exponential growth. The real part $g'$ induces an amplitude-dependent frequency shift, which will reduce the amplitude in the final result in the case of finite negative $g''$ or alternating exponential growth and decay in the case of a purely real $g$. 

\begin{figure}[h]
\includegraphics[]{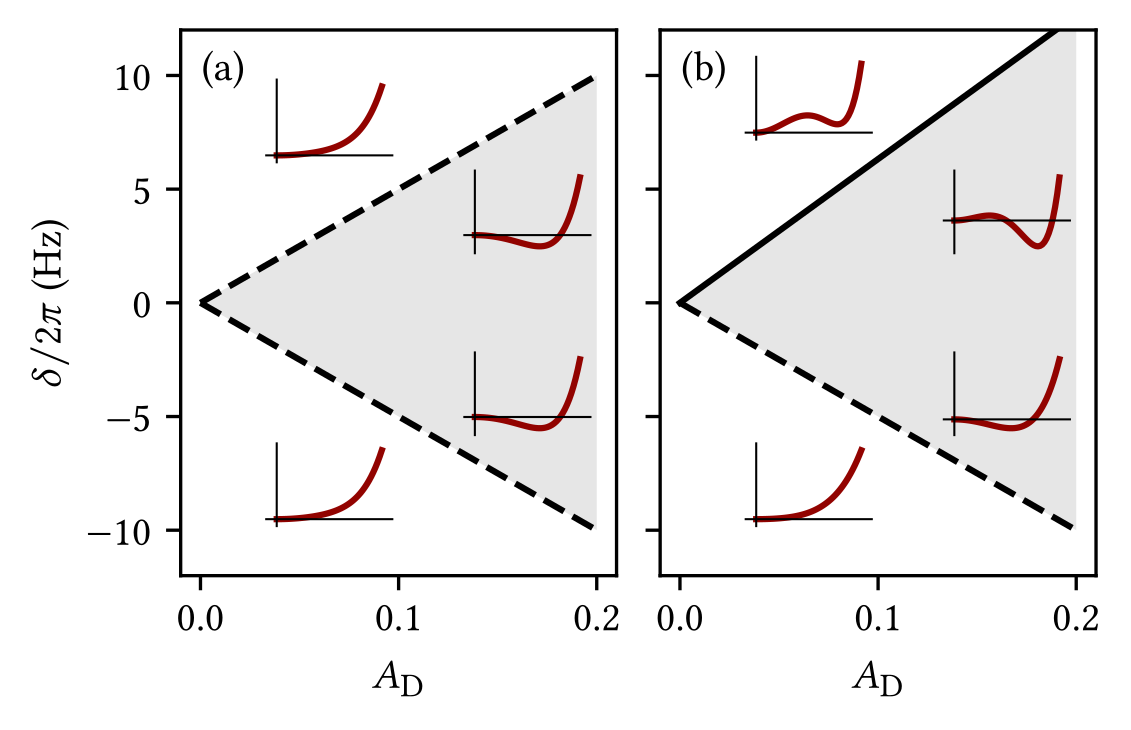}
\caption{Phase diagram as a function of driving amplitude and detuning, associated with the potential in Eq. (\ref{eq:pot_absa}) for \emph{(a)} $g'=0\,\mathrm{s}^{-1}$ and \emph{(b)} $g'=10^{-4}\,\mathrm{s}^{-1}$. The other parameters used are $\omgD = 2\pi\times 200\,\textrm{Hz}$ and $g'' = -10^{-4}\,\mathrm{s}^{-1}$. The shaded area indicates the parameters for which a stable space-time crystal is possible. The striped lines indicate smooth tricritical (left) or second-order (right) phase transitions between the crystalline and non-crystalline state. The solid line indicates a discontinuous first-order phase transition between the crystalline and non-crystalline state. The insets are a schematic representation of the potential of Eq. (\ref{eq:pot_absa}) in each region, plotted as a function of $|a|$. \label{phase_diagram}}
\end{figure}

Figure \ref{phase_diagram} shows the phase diagram associated with the potential in Eq. (\ref{eq:pot_absa}), where only the shaded area supports the existence of a stable space-time crystal. For $g' = 0$ shown in Fig. \ref{phase_diagram}a the phase transition between the non-crystalline and the time-crystalline phase is always of tricritical nature (smooth) and the transition point is found at $|\delta| = \frac{\omgD\AD}{4}$. For $g'>0$ shown in Fig. \ref{phase_diagram}b the graph is unchanged for $\delta<0$. However, for $\delta>0$ the phase transition becomes a (discontinuous) first-order phase transition and the line is shifted such that now%
\begin{equation}
\delta = \left(1-\frac{3}{4}\frac{(g')^2}{|g|^2} \right)^{-1/2}\frac{\omgD\AD}{4}.
\end{equation}%
As opposed to the case $g'=0$, for non-zero $g'$ and $\delta>0$ the minimum value for $|a|$ is finite up to the phase transition. This implies that a change in the  parameters such as variation of the detuning $\delta$ due to oscillations of the length of the condensate promptly causes the crystalline state to collapse.



\section{Experimental results \label{sc:exp_res}}
 
One of the challenges of the experiment is to excite the periodic (Floquet) drive sufficiently without inducing unwanted excitations. The condensate is trapped in an harmonic trap and in principle the drive, which is the radial breathing mode in our case, can be solely excited by modulating the radial trap frequency quickly. In practice, there are many complications to take into account. First of all, the trap is not purely harmonic and higher-order terms, although small, play a role. Furthermore, there are small imperfections in the construction of the trap, which lead to coupling between different modes in the trap. So apart from inducing the radial breathing mode, there are dipole modes in the radial and axial direction, a breathing mode in the axial direction and a scissor mode, which are induced simultaneously. In order to minimize the effects of other modes the current through the cloverleaf coils is reduced in a zigzag manner with a modulation depth of $\approx$ 10\% and returned to its original value, within a period of 10\,ms. This procedure is repeated a second time, such that the modulation can be kept small. This procedure minimizes the breathing mode in the axial direction, which is beneficial in the analysis of the experiment, since the crystalline mode that is excited depends strongly on the length of the condensate.

\begin{figure}[h]
\includegraphics{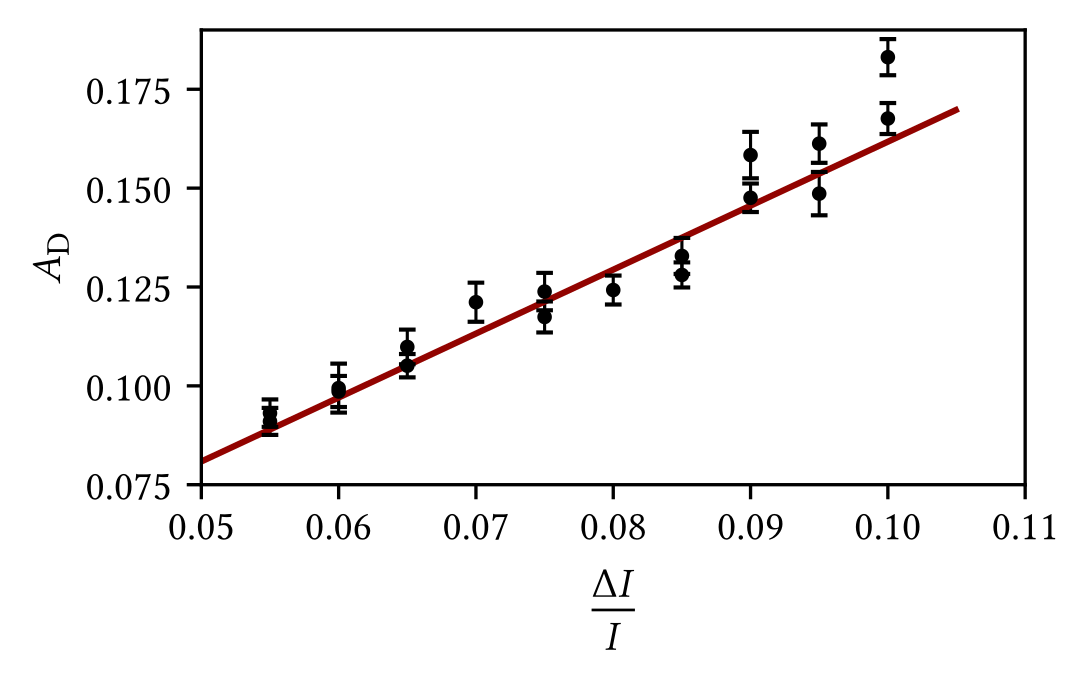}
\caption{The amplitude of the radial breathing mode for different modulation depths $\Delta I/I$ of the cloverleaf current. The amplitude of the breathing mode is  1.572 $\pm$ 0.017 times the modulation depth.  \label{fg:kick}}
\end{figure}

In Fig.~\ref{fg:kick} the amplitude of the breathing mode is plotted against the modulation depth, {\it i.e.}, the relative change $\Delta I/I$ in the current through the cloverleaf coils. The relative driving amplitude $\AD$ is directly proportional to the modulation depth and for the largest modulation depth $\AD$ becomes $\approx 0.17$. Although we can produce modulation depths that are much smaller than a minimum modulation depth of 0.05 in the experiment, the crystalline mode grows exponentially with a growth rate proportional to $\AD$ and small growth rates produce a detectable crystalline mode that grow in times longer than 1\,s.  Therefore we have used a minimum modulation depth of 0.05. At the same time the relative amplitude of the axial breathing mode remains below 2.5\%, which allows for single mode excitation of the crystal.

For the analysis of the mode induced we first fit the image using the column density of an unperturbed condensate. This fixes the position, length, width, number of particles and angle of the condensate in the images. The number of particles is about $17\times 10^6$, the length  is 120 pixels and the width is 7 pixels, where 1 pixel in the image corresponds to 2.43\,$\mu$m. Next we add a column density $\AX L_j(\bar{z})$ of the crystalline mode to the column density and fit the amplitude $\AX$ leaving the other parameters free. In the fit the mode number $j$ is varied and the proper mode $j$ is determined using several criteria. First of all, the mode that has the smallest $\chi_{\rm red}$ and the largest $\AX$ is selected. In all cases we either find a symmetrical, even mode $j=40$, or an asymmetrical, uneven mode $j=41$. Furthermore, it is checked that the position of the condensate in the axial direction is not changed appreciable. For example, an uneven mode has zero density in the center. If the crystalline mode increases the density in the center of the condensate, the fit procedure might shift the condensate to the side in order to minimize $\chi_{\rm red}$, indicating that the proper mode in that case is not uneven. Finally, it is checked that the amplitude $\AX$ displays a sinusoidal oscillation in time, since shifting the condensate might change the sign of the fitted $\AX$. Apart from the two modes indicated, no other modes have been detected. It shows that our imaging system has sufficient resolution to distinguish between two adjacent modes, and that our excitation scheme is tailored to excite only one single mode.

The holographic imaging technique allows for the determination of the column density $\rho_c$ of the atomic cloud and due to the cylindrical symmetry of the cloud for the determination of the density $\rho$. Since the crystalline mode has no radial dependence and the mode only exists within the condensate, the density ${\rho}$ of the  crystalline mode is given by $\rho = \rho_c/(2 R_\rho)$ with $R_\rho$ the radial width of the condensate. As shown in Ref.~\cite{liao19} the density of a crystalline mode containing one quantum is given by
\begin{equation}
q_j = \sqrt{\frac{\hbar \omgD}{2 \TtwoB R_\rho{}^2 R_z Q_{jj}}}, \label{eq:dens}
\end{equation}
where the factor $Q_{jj}$ is a known normalization constant for the overlap of the mode within the condensate. Equation~\ref{eq:dens} allows us to express the amplitude $\AX$ in terms of the number of quanta in the crystalline mode.

\subsection{Breaking discrete time symmetry\label{sc:breaking}}

\begin{figure}[h]
\includegraphics{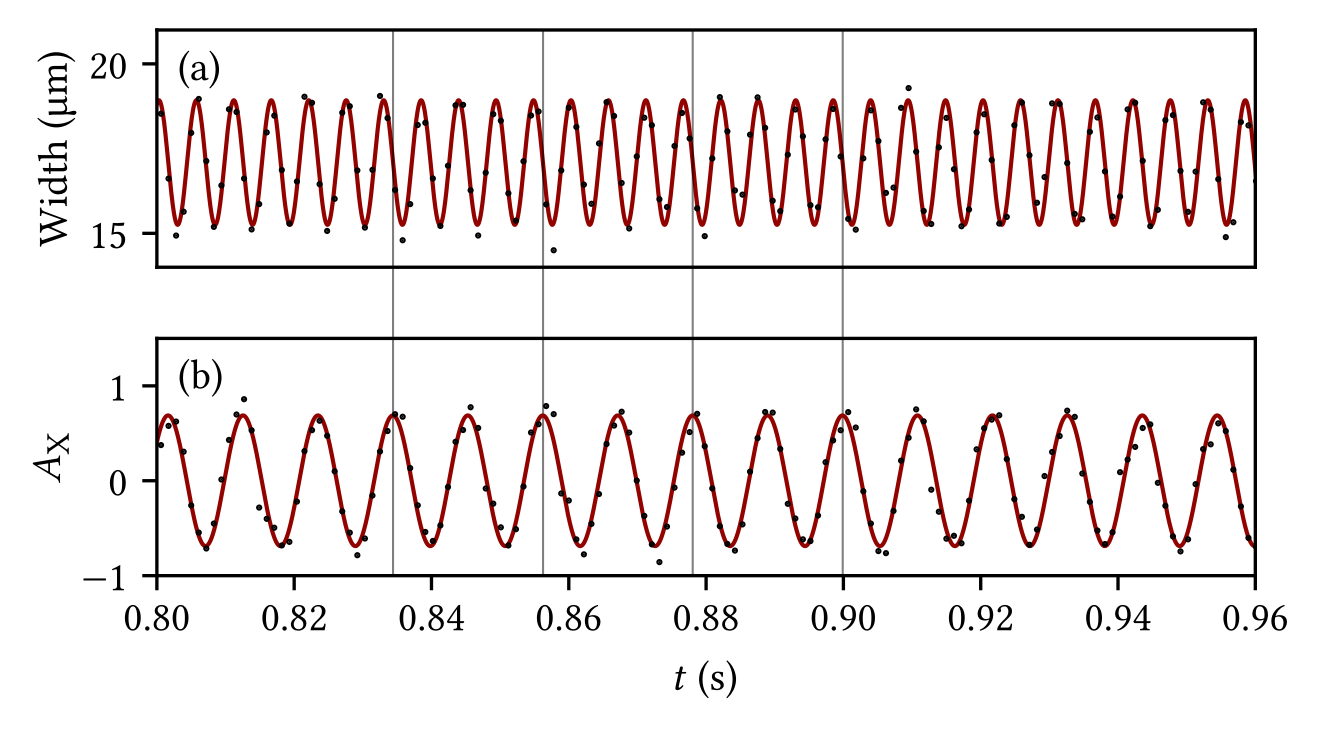}
\caption{The radial breathing mode {\it vs.} the crystalline mode for a modulation depth of 0.0675. Here the width of the condensate is given in pixels, where 1 pixel corresponds to 2.43\,$\mu$m and the radial breathing mode has a frequency of 183.26 $\pm$ 0.04\,Hz. The crystalline mode is $j=40$ and oscillates at a frequency of 91.61 $\pm$ 0.03\,Hz, which is within the uncertainty a factor 2 smaller. \label{fg:focus_freqs}}
\end{figure}

In Fig.~\ref{fg:focus_freqs} the amplitude of the breathing mode in the radial direction is shown together with the amplitude $\AX$ of the crystalline mode. From the figure it is immediately clear that the discrete time-symmetry of the periodic drive, the radial breathing mode, is broken and that the crystal oscillates with half the frequency. Note that the amplitudes of both modes are constant and that the crystalline mode is in phase with the breathing mode, although the breathing mode has been excited 0.75 s or nearly 140 periods before. It shows the small damping and large coherence in our superfluid system, where the fraction of thermal atoms is very low ($< 10\%$). 

\subsection{Growth rate \label{sc:growth}}

\begin{figure}[h]
\includegraphics{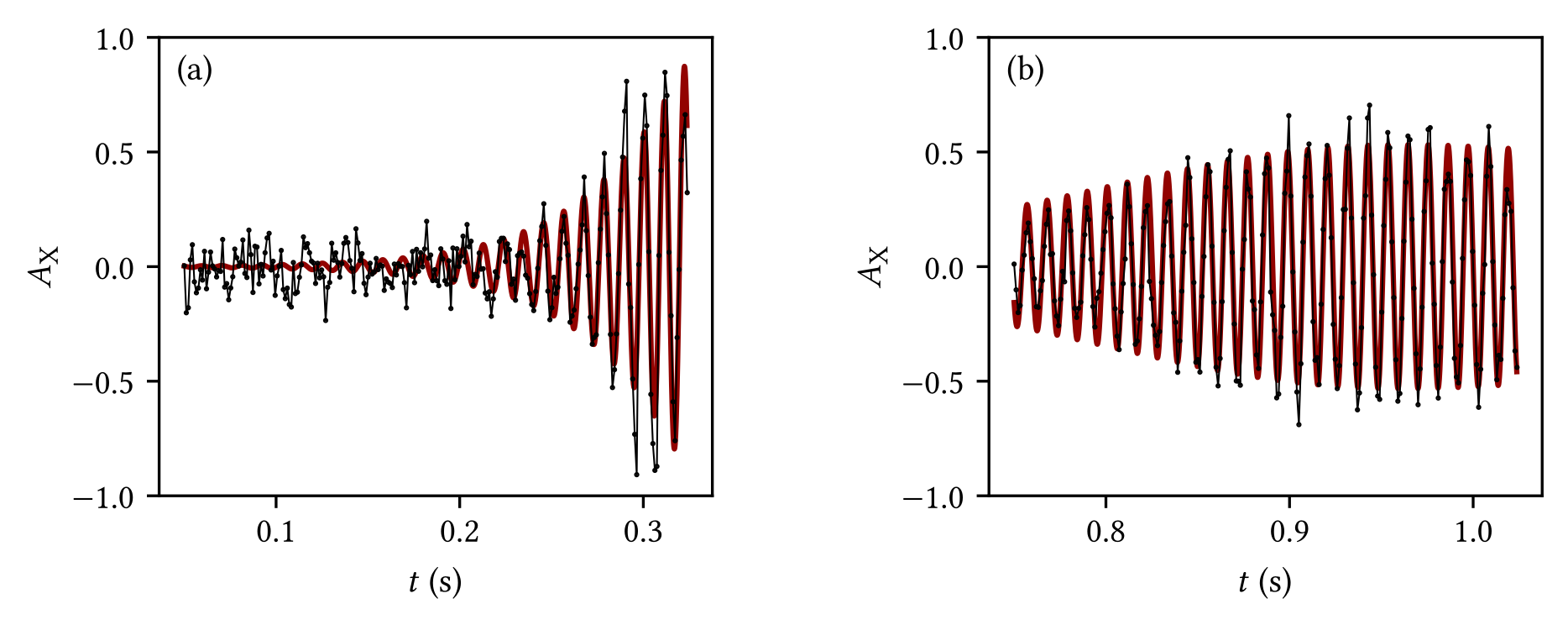} %
\caption{Initial growth rate of the space-time crystal pattern. The points represent the amplitude of then mode with $j=41$ fitted to the data as in Figure \ref{profile}. The line is the model of Eqs. (\ref{eq:a_eom}) and (\ref{eq:as_eom}) with fixed starting conditions $a=1$, $\delta=3.07\,\textrm{Hz}$, and $g$, the frequency $\omgX$ and phase as fit parameters. \emph{(a)} Observation of the growth right after a strong excitation. The space-time crystal appears after approximately 0.2\,s. Fitting the model to the data with fixed $\AD=0.1$ results in $|g|=0.19(25)\times10^{-4}\,\textrm{s}^{-1}$. \emph{(b)} Observation of the growth a long time after a weaker excitation. The pattern is still growing after 0.75\,s. Fitting the model to the data with fixed $\AD=0.072$ results in $|g|=0.404(2)\times10^{-4}\,\textrm{s}^{-1}$.  \label{kick}}
\end{figure}

In the experiments the crystalline amplitude $\AX$ is detected for various modulation depths after a hold time of 0.05 and 0.75\,s. In the first case, the crystalline mode is only observed for the highest modulation depth of 0.1, whereas in the other case the crystal has already saturated to the final value for most of the modulation depths. Since the density of the crystal is superposed on the density of the condensate and its amplitude is small, the crystal grows quickly from below the detection threshold to its limiting value. Thus we can only observe the growth of the crystal during a small number of oscillations. The result for a hold time of 0.05\,s is shown in Fig. \ref{kick}a. The space-time crystal starts growing above the detection threshold about $0.2\,\textrm{s}$ after the excitation. To model the observations it is assumed that the field $a=1$ at $t=0$ and the equations of motion from Eqs. (\ref{eq:a_eom}) and (\ref{eq:as_eom}) are used to model the growth using only the detuning $\delta$ and $g$ as free parameters. For the driving amplitude $\AD$ as determined from the modulation depth we find $\delta=3.07\,\textrm{Hz}$, indicating that the growth rate is strongly reduced due to the detuning $\delta$. The optimal value of $g$ has a large uncertainty, since the signal is not yet saturated at the end of the measurement. 

For a larger holding time of 0.75\,s the growth is close to saturation, as shown in Fig. \ref{kick}b. The pattern is shown to grow until $t\approx 0.9\,\textrm{s}$, when the pattern saturates and levels out. Again the model is fitted using only $\delta$ and $g$ as free parameters. Now $|g|$ is accurately determined with only half of a percent relative uncertainty. It is seen that the model, despite having only a few free parameters, fits the data well. Note that in the model we have assumed that the dominant higher-order term allowed by symmetry is of fourth order, whereas in principle terms of any even order are possible. Thus the value of $g$ found from the model does not have to correspond exactly to the fourth-order term from the model, since higher-order terms in the Hamiltonian might also contribute to the saturation.

\subsection{Saturation \label{sc:stability}}
One of the key issues in the discussion about time crystals is the long-term stability. In some cases it is argued that the many-body localization plays a crucial role in the stabilization process, whereas in other cases it is shown theoretically that crystals can be stabilized in a so-called pre-thermal phase. Since our system can be described nearly ab-initio and a direct comparison between model and experiment is feasible, we can test the requirements for the long-term stability of  the space-time crystal in our case. By adding the fourth-order term to the Hamiltonian, the number of quanta in the crystalline mode stabilizes as given by Eq.~(\ref{eq:eq_absa}). Note that the imaginary part of $g$ provides for damping, whereas the number of quanta becomes purely oscillatory in case the imaginary part of $g$ is exactly zero.  

\begin{figure}[h]
\includegraphics{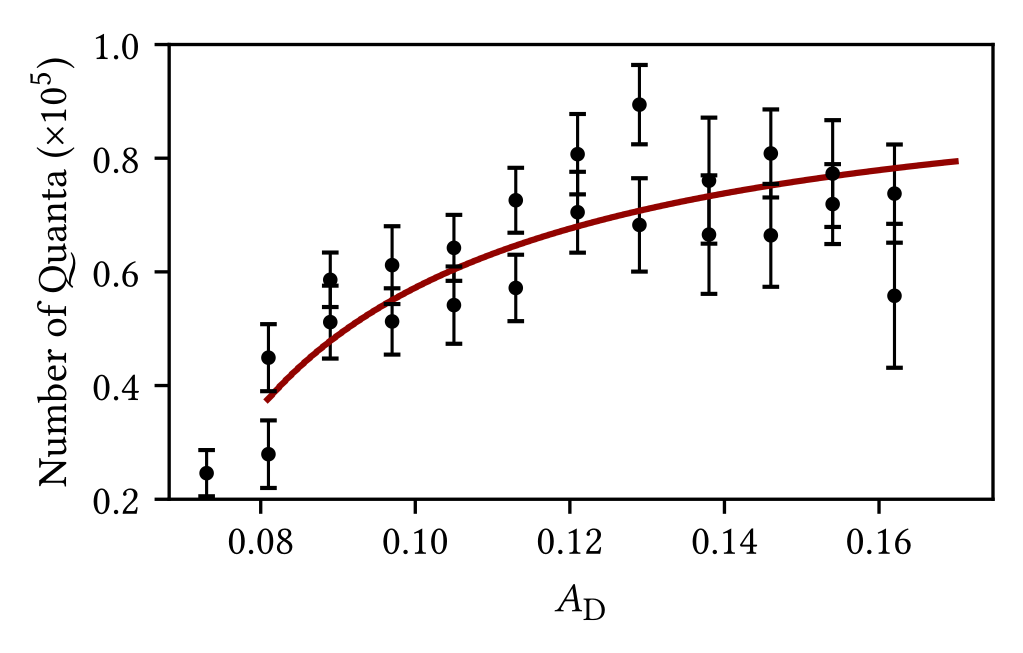}
\caption{The number of quanta in the space-time crystal at saturation, for different drive strengths. The line indicates the maximum possible number of quanta for each drive strength, derived from Eq. (\ref{eq:eq_absa}) for fixed $\delta = 3.04\,\textrm{Hz}$ and fitted $|g|/\AD=0.48(5)\times10^{-3}\,\textrm{s}^{-1}$. \label{stability}}
\end{figure}

To study the saturation of the number of quanta in the space-time crystal mode, the number of quanta in this mode is studied after $0.75$\,s for those cases, where the modulation depth is sufficient for the number of quanta to stabilize. The results are shown in Fig.~\ref{stability}. As can be seen from the figure, the number of quanta slightly increases for higher modulation depth. This indicates that for the modulation depths used in the experiment the absolute value of the detuning $|\delta|$ is smaller than the driving term $\AD \omgD/4$ as expected. The expression of Eq. (\ref{eq:eq_absa}) is fitted to the data assuming $|g|$ proportional to $\AD$, which is reasonable given the fact that the fourth-order terms contributing to $g$ are proportional to the deformation of the condensate. The fits yields a value of $\delta=3.04\,\textrm{Hz}$ and the value of $g$ is found to be in close agreement with the fit to the data in Fig.~\ref{kick}. Thus all experimental data can be explained using the model described in Sec.~\ref{sc:modelling} using similar parameters and this lends further credibility to our model. 

\section{Conclusion and outlook}
By deforming the trap during the final stages of the cooling process it is possible to create a Bose-Einstein condensate with large aspect ratio. Radially exciting this BEC realizes a stable Floquet space-time crystal, in which the radial excitation acts as a periodic drive, while the space-time crystal is a higher-order axial mode. Using a minimally destructive holographic imaging method the dynamics of this space-time crystal is studied for 250 recordings or 30 periods of the space-time crystal. By calculating a line-density profile for each recording and studying the behavior of these line-density profiles in time, a pattern periodic in space and time is observed. By calculating the Fourier transform of a time-series of these line-density profiles, a quartet of peaks is observed in a rectangular orientation. The peaks are no more than a pixel wide along the frequency axis, indicating coherence over the entire time of the recording. The rectangular orientation of the peaks is a clear indication of long-range space-time crystalline ordering. The frequency at which the peaks appear corresponds to half the driving frequency, which is confirmed  within margin of uncertainty by direct fitting of the crystal amplitude.

To better understand the saturation of the crystal observed in experiments in this work and previous works the original model is expanded with a fourth-order term which has previously been neglected. It is found that addition of this complex-valued fourth-order parameter explains the saturation. Moreover, this fourth-order term leads to an interesting phase diagram exhibiting a tricritical point and both smooth and discontinuous phase transitions.

It is shown that the excitation of the driving mode is in linear relation with the modulation of current through the coils generating the trapping field. By carefully modulating only the coils responsible for the radial trapping frequency the axial size, or length, of the BEC is not disturbed, leading to an oscillation only along the radial direction. The emergent space-time crystal appears at exactly half the frequency of the driving mode and is phase-locked to the driving mode, breaking discrete time translation symmetry. Using the fourth-order model, the initial growth rate and saturation of the space-time crystal can be adequately described assuming that the fourth-order coefficient is proportional to the amplitude of the driving mode. The model is also used to calculate the saturation amplitude of the space-time crystal for a range of driving mode amplitudes. Good agreement between parameters in the model is found when comparing the growth-rate and saturation data, showing the efficacy of the presented model.

Our system provides an excellent platform for studying space-time crystalline phenomena as the space-time crystal can be very well modeled using a nearly {\it ab-initio} model. Minimally destructive imaging can be used to study perturbations of the crystalline phase for an extended period of time. Future work can be directed towards exploring the presented phase diagram, which exhibits a discontinuous phase transition for positive detuning which might support meta-stable space-time crystalline states and therefore hysteresis close to the transition line. It is also possible to introduce a controlled excitation of the length of the condensate, the effect of which should be adequately described using the phase diagram and which might lead to the space-time crystalline ordering appearing and reappearing multiple times over prolonged period of time.


\section*{Acknowledgments}
This work is supported the Stichting voor Fundamenteel Onderzoek der Materie (FOM) and is part of the D-ITP consortium, a program of the Netherlands Organization for Scientific Research (NWO) that is funded by the Dutch Ministry of Education, Culture and Science (OCW).

\bibliographystyle{apsrev}
\bibliography{StabilitySTXtalNotes}

\end{document}